\documentclass[a4paper,11pt,oneside,openany]{article}
\usepackage{graphicx}
\usepackage{wrapfig}
\usepackage{url}
\usepackage{comment}
\usepackage[top=30truemm,bottom=30truemm,left=30truemm,right=30truemm]{geometry}
\usepackage{amsmath}
\usepackage{authblk}
\usepackage[left]{lineno}

\title{New high-precision measurement system for electron-positron pairs from sub-GeV/GeV gamma-rays in the emulsion telescope}

\author[1,2,*]{Yuya Nakamura}\author[3]{Shigeki Aoki}\author[1]{Tomohiro Hayakawa,}\author[4]{Atsushi Iyono}\author[3]{Ayaka Karasuno}\author[5]{Kohichi Kodama}\author[1]{Ryosuke Komatani}\author[1]{Masahiro Komatsu}\author[1]{Masahiro Komiyama}\author[3]{Kenji Kuretsubo}\author[3]{Toshitsugu Marushima}\author[3]{Syota Matsuda}\author[1]{Kunihiro Morishima}\author[1]{Misaki Morishita}\author[1]{Naotaka Naganawa}\author[1]{Mitsuhiro Nakamura}\author[3]{Motoya Nakamura}\author[3]{Takafumi Nakamura}\author[1]{Noboru Nakano}\author[1,2]{Toshiyuki Nakano}\author[1]{Akira Nishio}\author[3]{Miyuki Oda}\author[1]{Hiroki Rokujo}\author[1]{Osamu Sato}\author[1]{Kou Sugimura}\author[3]{Atsumu Suzuki}\author[3]{Satoru Takahashi}\author[1]{Mayu Torii}\author[4]{Saya Yamamoto}\author[6]{Masahiro Yoshimoto}

\affil[1]{Nagoya University, Nagoya 464-8602, Japan}
\affil[2]{Kobayashi-Maskawa Institute for the Origin of Particles and the Universe, Nagoya University, Nagoya 464-8602, Japan}
\affil[3]{Kobe University, Kobe 657-8501, Japan}
\affil[4]{Okayama University of Science, Okayama 700-0005, Japan}
\affil[5]{Aichi University of Education, Kariya 448-8542, Japan}
\affil[6]{Gifu University, Gifu 501-1193, Japan}
\affil[*]{E-mail: ynakamura@flab.phys.nagoya-u.ac.jp}

\date{}

\begin{document}

\maketitle

\begin{abstract}
{The GRAINE project observes cosmic gamma-rays, using a balloon-borne emulsion-film-based telescope in the sub-GeV/GeV energy band. We reported in our previous balloon experiment in 2018, GRAINE2018, the detection of the known brightest source, Vela pulsar, with the highest angular resolution ever reported in an energy range of $>$80\,MeV. However, the emulsion scanning system used in the experiment was designed to achieve a high-speed scanning, and it was not accurate enough to ensure the optimum spatial resolution of the emulsion film and limited the performance. Here, we report a new high-precision scanning system that can be used to greatly improve the observation result of GRAINE2018  and also be employed in future experiments. The scanning system involves a new algorithm that recognizes each silver grain on an emulsion film and is capable of measuring tracks with a positional resolution for the passing points of tracks of almost the same as the intrinsic resolution of nuclear emulsion film\ ($\sim$70\,nm). This resolution is approximately one order of magnitude smaller than that obtained with the high-speed scanning system. With this scanning system, an angular resolution for gamma-rays of 0.1$^\circ$ at 1\,GeV is expected to be achieved. Furthermore, we successfully combine the new high-precision scanning system with the existing high-speed scanning system, enabling the high-speed and high-precision measurements. Employing these techniques, we reanalyze the gamma-ray events detected previously by only the high-speed scanning system in GRAINE2018 and obtain an about three times higher angular resolution\ (0.22\,$^\circ$) in the 500--700\,MeV energy range. Adopting this technique in future observations may provide new insights into the gamma-ray emission from the Galactic center region and may realize polarization measurements of high-energy cosmic gamma-rays.}
\end{abstract}

\newpage
\tableofcontents
\newpage

\section{Introduction}
\label{sec:intro}

The observation of cosmic gamma rays is crucial for understanding high-energy astrophysical phenomena and the mechanism of cosmic-ray acceleration. The most significant recent progress in the field was achieved with the Large Area Telescope on the Fermi Gamma-ray Space Telescope (Fermi-LAT) launched in 2008. It observes the sub-GeV/GeV gamma-ray sky\cite{fermi}\cite{fermiCatalog}, yielding many significant results. One of the highlights is the discovery of cosmic-ray proton acceleration in supernova remnants, derived from the gamma-ray emission morphology and spectrum\cite{W44}. Its acceleration mechanism remains yet unknown, though, mainly because the relatively-poor angular resolution of the Fermi-LAT limits the precision of morphological comparisons with observation results in other wavelengths to determine the detailed emission region. The Fermi-LAT also detected unexpected GeV gamma-ray emission from the Galactic center region and the emission may be from the annihilation of dark matters \cite{GeVexcess}. Though it failed to identify the origin sources due to its insufficient angular resolution to resolve potential sources located in a diffuse gamma-ray background. Thus, observations with a higher angular resolution are  essential to resolve these problems and advances sub-GeV/GeV gamma-ray astronomy, which still is in its infancy majorly because of the technical difficulty in observation.

The Gamma-Ray Astro-Imager with Nuclear Emulsion (GRAINE) project aims to observe cosmic gamma rays with high-angular resolution and polarization sensitivity in the 10\,MeV--100\,GeV band using a balloon-borne telescope equipped with a nuclear emulsion chamber \cite{GRAINE}. The nuclear emulsion chamber encompasses about a hundred sheets of stacked emulsion films in its core. The nuclear emulsion film is a three-dimensional tracking detector with a sub-micron spatial resolution. Each film composed of a plastic base film and emulsion layers applied on both sides of the base film, and we produce the films by ourself in Nagoya University\cite{RtR}. AgBr crystals are dispersed in the emulsion layers with the crystal volume occupancy of about 40\%, and they have sensitivity for charged particles. Each crystal that charged particles penetrated generate silver grain after the photo-development, and a track of a charged particle is recorded in a form of a series of silver grains, each of which has a diameter of $\sim$1\,$\mu$m. These analogically recorded tracks were scanned by eye using microscope and this time-consuming method limited the statistics of the analyzable events. In the past 25 years, we have recognized tracks by an automatic emulsion scanning system. In the system, objective lens and imaging camera take the tomographic images of emulsion layers, and processing computer digitally process the images and create digital track data in a short time. The nuclear emulsion film can precisely measure the angles of the electron and positron tracks produced in pair production of gamma-rays ($\gamma$ (Z or e$^-$)\,$\rightarrow$\,e$^+$e$^-$  (Z or e$^-$), where Z  denotes a nucleus) near the conversion point. Thus, the angular resolutions obtained with the nuclear emulsion film is very high for gamma rays (1$^\circ$ at 100\,MeV and 0.1$^\circ$ at 1\,GeV), which are 5-8 times higher than that with the Fermi-LAT\ (5$^\circ$ at 100\,MeV and 0.8$^\circ$ at 1\,GeV\cite{FermiAngRes}). Furthermore, whereas the Fermi-LAT and the other experiments have not detected the polarization of cosmic gamma-rays in this energy range yet, the emulsion film has high sensitivity for the polarization and we verified the performance in the polarized GeV gamma-ray beam experiment by decoding the precisely observed azimuthal-angle distribution of the produced pairs of electrons and positrons, which carries information of the polarizations of incident gamma-rays\cite{GRAINE_polar}. The polarizations have the information of the magnetic field at the emission region, which is important for investigating the cosmic-ray acceleration mechanism. The GRAINE project plans to perform repeated observations with an aperture area of 10\,m$^2$ with the flight duration of one week. A primary target is the very central region\ ($<$0.1$^\circ$) in the Galactic center. The high-angular resolution realizes the observation with very low contamination of the diffuse gamma-ray background (8 times higher angular resolution makes 1/64 contamination of the diffuse gamma-ray), and it will help us to reveal the emission source(s) by comparing it with expected galactic sources, assumed dark-matter density profiles, and so on. Other important classes of targets include the supernova remnant, pulsar and blazar; precise morphological studies of supernova remnants and/or polarization measurements of supernova remnants/pulsars/blazars will provide unprecedented insights into the cosmic-ray acceleration mechanism.

In the GRAINE project, detectors have been developed and several balloon experiments have been conducted to study the performance of the emulsion gamma-ray telescope in astronomical observations\cite{GRAINE_polar}\cite{GRAINE2011}\cite{GRAINE2015}. With the detector, a gamma-ray event detection begins with a search for two closely-situated tracks produced in the stacked emulsion films, followed by reconstruction of the energy and incidence-angle of the identified candidate gamma-ray from the tracks. Since the initial interaction, a pair production from a gamma-ray, in the detector can happen anywhere in the detector and since there is no way to know where it happens a priori, all charged-particle tracks recorded on all the films, or as many of them as possible, should be read out and digital track data is made by the emulsion scanning system for searching. Also, the scanning speed is one of the key parameters, given that the total area of the films is huge for hundreds of layers of emulsion sheets in the large aperture area chamber. A team in Nagoya University, including some of the authors of the present paper, developed a high-speed scanning method for it; the method takes some discrete tomographic images of the emulsion layers, applying binarization, and searches for an aligned series of hit pixels\cite{TS}. Our third balloon experiment, GRAINE2018, was performed in Australia in 2018 with an aperture area of 0.38\,m$^2$ with the aim of detecting the brightest known gamma-ray source, the Vela pulsar, as a performance verification of the integrated system, in which the most up-to-date high-speed scanning system at the time, Hyper Track Selector (HTS), was employed\cite{HTS}. GRAINE2018 clearly detected the Vela pulsar with an angular resolution (68\%-events containment radius) of 0.42$^\circ$ for energies above 80\,MeV (mainly 100--700\,MeV). We realized the large-scale experiment with the nuclear emulsion with the high-speed scanning system, and this result was the first real proof of a working emulsion gamma-ray telescope that has the highest angular resolution among telescopes in the sub-GeV energy band \cite{GRAINE2018_1}\cite{GRAINE2018_2}. 

Although GRAINE2018 achieved the highest angular resolution in the sub-GeV band, there remains a great deal of margin for improvement, given that the resolution was limited mainly by the measurement accuracy of the high-speed scanning system. In the simple model that takes into account only the intrinsic positional resolution of the analogically recorded track in the nuclear emulsion film ($\sim$0.07\,$\mu$m) and effect of multiple Coulomb scattering in the film, the expected (ideal) angular resolution is $\sim$0.25$^\circ$ in GRAINE2018 for 100--700\,MeV; the obtained angular resolution (0.42$^\circ$) is larger than this expected value, because the positional resolution realistically depends on the measurement accuracy of the emulsion scanning system which makes digital track data from analogically recorded tracks. The measurement accuracy of the HTS depends on the pixel size of the tomographic image, 0.45\,$\mu$m, and the distance between each tomographic image, $\sim$4\,$\mu$m; the expected measurement accuracy of the incident angle of the gamma-ray measured by the HTS without considering the effect of multiple Coulomb scattering is $\sim$0.13$^\circ$ at tan$\theta_\gamma$=0.1 and $\sim$0.44$^\circ$ at tan$\theta_\gamma$=1.0 (here,\ tan$\theta_\gamma$ is the incidence-angle of the gamma-ray and the direction of tan$\theta_\gamma$ = 0 is set nearly equal to the direction of the zenith). One of the main goals of the GRAINE project for high-angular-resolution GeV observation is to resolve the GeV gamma-ray emission from the Galactic center region, and it requires an angular resolution of 0.1$^\circ$ at 1\,GeV. However, we can not realize it with the HTS due to its low measurement accuracy. High-energy gamma-ray observation with small effect of multiple Coulomb scattering tends to be particularly prone in quality to a large effect of measurement accuracy of the scanning system. Furthermore, the Vela pulsar was observed with 0.3$<$tan$\theta_\gamma$$<$1.0 in GRAINE2018, and the large angular dependence of the measurement accuracy significantly affects the observation performance.

We develop a new emulsion scanning system with a high measurement accuracy to keep the intrinsic performance of the emulsion film. It is referred to as the "high-precision scanning system". We develop a new algorithm to measure three-dimensional position of each silver grain of the recorded track to improve the measurement accuracy because the discrete information of hit pixels limit the performance in the high-speed system. Because the emulsion films analogically record tracks, we can repeatedly rescan and reanalyze with a different scanning system. Then, since the scanning speed of the high-precision system with the new algorithm is slow, only the small area of the emulsion film in the vicinity of the recorded gamma-ray events is rescanned. It means that the new system delegates to the high-speed system the job of searching for gamma-ray events from a large number of tracks and reanalyzes the selected events only to much higher precision. This new method realize the measurement with both of high-speed and high-precision. 

In this paper, we describe the new high-precision scanning system and also a performance evaluation with it. We use flight films in GRAINE2018 for developing it and we present an overview of GRAINE2018 in Section 2. In section 3, we describe our high-precision measurement system and new algorithm, and present its basic performance, including the position measurement accuracy and angular resolution for charged-particle tracks. In section 4, the performance for gamma-ray events is demonstrated, using the flight data in GRAINE2018. Summary of the results and future prospects are presented in Section 5. A preliminary study of this work was presented in \cite{ICRC}.

\section{GRAINE2018 experiment}
\begin{figure}
\center
\includegraphics[bb=0 0 1289 833, width=.6\textwidth]{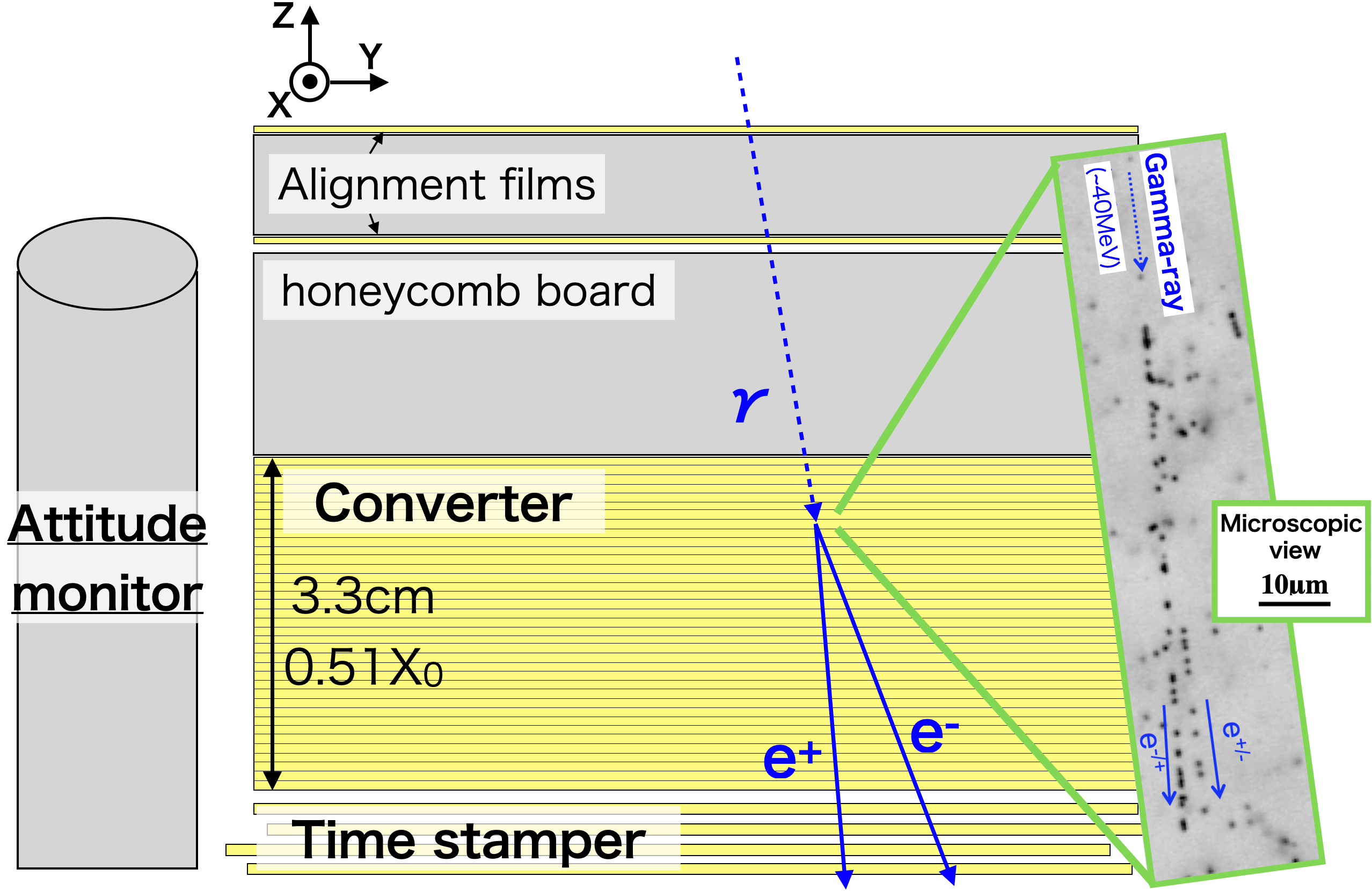}
\caption{Cross-sectional view of the emulsion gamma-ray telescope, consisting of an attitude monitor (on the left) and emulsion chamber (right). A photo of a microscopic view of a pair production event is also displayed as an inset.}
\label{fig:telescope}
\center
\end{figure}

The emulsion gamma-ray telescope consists of an emulsion chamber and a star camera for the attitude monitor (Figure \ref{fig:telescope}). Incoming gamma rays produce electrons and positrons in the converter in the emulsion chamber. In the GRAINE2018 configuration, the converter consisted of 100 emulsion films, each composed of a 180-$\mu$m-thick plastic base film and 75-$\mu$m-thick emulsion layers applied on both sides (measuring 25 $\times$ 38\,cm$^2$), and we used 4 converters. The diameter of each piece of AgBr crystal, which has sensitivity for charged particles, in the emulsion layers is about 240\,nm. All tracks are recorded in a form of a series of silver grains after a development process (see the inset in Figure \ref{fig:telescope}). Each event is mapped with the celestial coordinates from the observed gamma-ray direction, time obtained with the multistage emulsion shifter \cite{shifter}, and attitude information obtained with the star camera. We place films on the top of the converters and these films vacuum-packed with a flat aluminum honeycomb board to maintain their flat shapes. These films are referred to as "alignment films" and these films are used as a reference of the flatness of the films in the converters. The relative angles between the alignment films and star camera was measured by the three dimensional coordinates measurement instrument, FaroArm. The GRAINE2018 balloon experiment was conducted on 2018 April 26 in Alice Springs, Australia. The flight lasted for approximately 17\,h and reached an altitude of more than 35\,km (with a residual atmosphere of\ 3--5\,g/cm$^2$). 

\section{High-precision scanning system}
\subsection{Hardware of the developed system}
\begin{figure}
\center
\includegraphics[bb=0 0 1066 817, width=.8\textwidth]{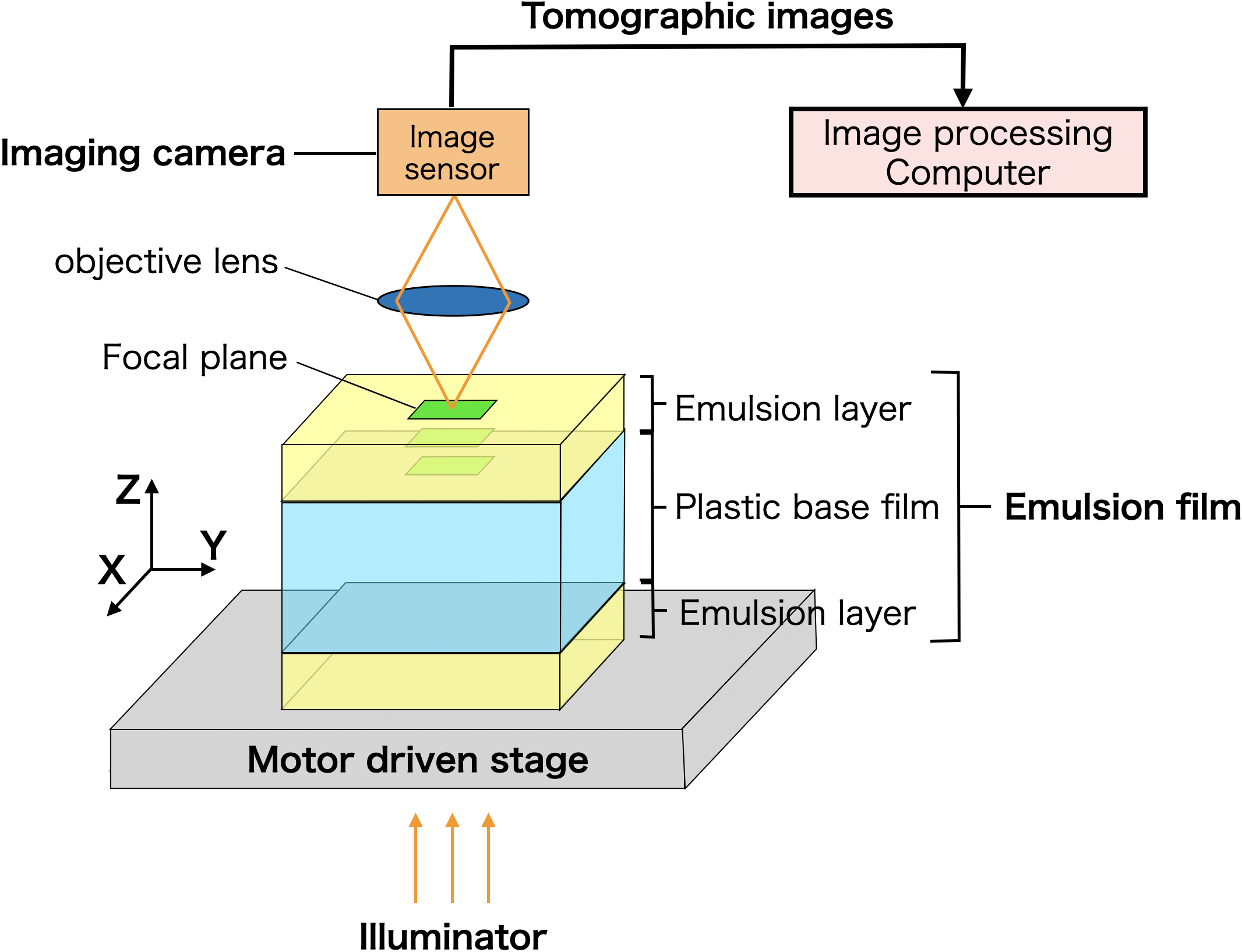}
\caption{Outline of image taking process in the emulsion scanning system.}
\label{fig:proc}
\center
\end{figure}

Figure \ref{fig:proc} illustrates the image-taking process in the emulsion scanning system. The main components of the system are an illuminator, objective lens, imaging camera, and motor-driven stage. The film is set on the motor-driven stage. Hereafter, the plane parallel on the film is defined as the X-Y plane, and the direction perpendicular to it is defined as the Z-axis. The imaging camera takes tomographic images in the emulsion layer, and motor-driven stage moves along the X-Y plane to take images in other view. The hardware of the new system is based on the previous emulsion-scanning system, called the Ultra Track Selector, which is an advanced model of the Track Selector \cite{TS}\cite{UTS}. This new system employs new camera and the image pixel size of this system is 0.15\,$\mu$m $\times$ 0.15\,$\mu$m, whereas that of the HTS is 0.45\,$\mu$m $\times$ 0.45\,$\mu$m. Table \ref{tab:lens} shows the specification of the objective lens and the numerical aperture is 0.85, and the illuminator is a halogen lamp with a green\ ($\sim$550\,nm) filter. The optical resolution in the system is calculated from these values to be 0.61$\times$550/0.85 $\simeq$ 395\,nm, whereas that of the HTS is 0.61$\times$436/0.65 $\simeq$ 409\,nm. This system has about 154\,$\mu$m $\times$ 154\,$\mu$m field of view. The electron and positron tracks of gamma-ray events are recorded within 75\,$\mu$m $\times$ 75\,$\mu$m region in each emulsion layer for tan$\theta_\gamma$<1.0, and the field of view of the high-precision system is enough size to take images only around the gamma-ray events. The HTS has a large field of view of 5.1\,mm $\times$ 5.1mm. Accordingly, it is capable of identifying many tracks in one view and is suitable for high-speed scanning.
\begin{table}
\centering
\scalebox{0.9}[0.9]{
\begin{tabular}{lll}
     & high-precision system & HTS \\ \hline \hline
    manufacture & TIYODA & KONICA MINOLTA \\
    magnification & 50$\times$ & 12.1$\times$ \\
    Numerical aperture & 0.85 & 0.65 \\ \hline
\end{tabular}}
\caption{Specification of the objective lens in our latest high-speed system, HTS, and the high-precision system.}
\label{tab:lens}
\end{table}

\subsection{Algorithm handling each silver grain that composes a track}
\begin{figure}
\center
\includegraphics[bb=0 0 1457 685, width=.9\textwidth]{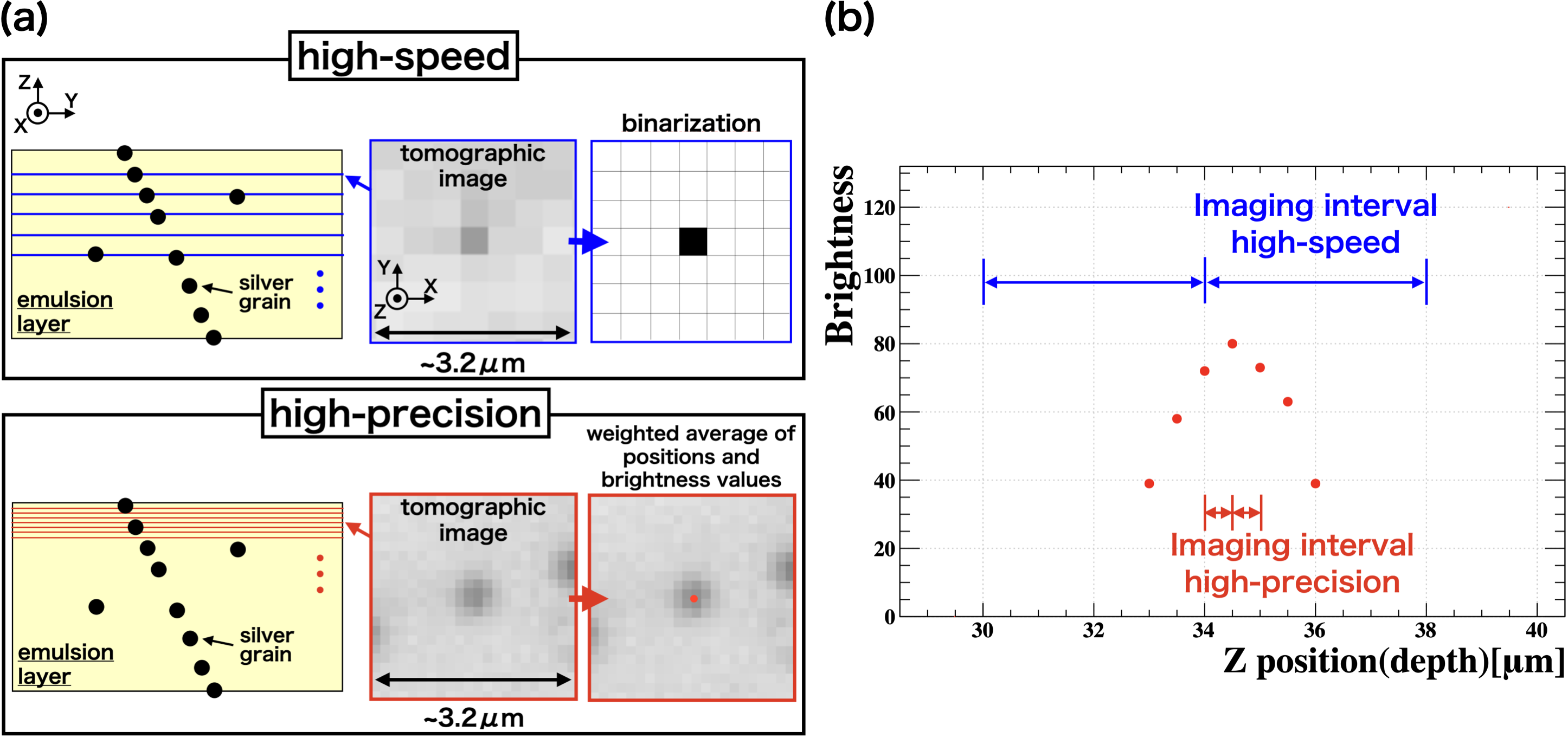}
\caption{(a): Schematic view of tomographic images of the emulsion layers in the high-speed and high-precision systems. The central panel in each row is a real close-up image for one silver grain. The events in the two rows are independent. (b): Depth dependence of the brightness of one grain taken in the high-precision system in red dots. Images are taken in an interval of 0.5\,$\mu$m with a total of 150 tomographic images in the high-precision system and in an interval of 4\,$\mu$m with 16 images in the high-speed system (blue lines with arrows simply demonstrating the interval) .}
\label{fig:scan}
\center
\end{figure}

We develop a new algorithm that identifies each silver grain recorded in the emulsion film for the measurement with a high positional resolution. Figure \ref{fig:scan} shows the differences in taking tomographic images and image-processing algorithm between those in the high-speed and in the high-precision system. Figure \ref{fig:scan}(a) shows a schematic view of the emulsion layer in the emulsion film and tomographic image of the emulsion layer taken by the high-speed scanning system and by the high-precision system, respectively. In the high-speed system, images are binarized, and tracks are searched through 16 binarized tomographic images, using only simple shift and sum functions, and the high-speed system detects tracks as simple straight lines\cite{TS}\cite{HTS}. In the high-precision system, by contrast, images are temporarily binarized, neighboring pixels are identified, and clustered pixels are labeled. The position on the X-Y plane of each silver grain is determined as the weighted average of the pixel positions and the brightness values stored in the pixels in each cluster. In addition, the Z position, thus the three-dimensional position, of each grain can be precisely determined as the weighted average of the Z-positions obtained in the many images in the high-precision system, in contrast to the high-speed system, where the obtained images are too few in number and too poor in quality to derive the Z position with the same precision. Figure \ref{fig:scan}(b) shows the obtained depth-dependence of the brightness of one grain taken in the high-precision system, where a total of 150 tomographic images are taken with an interval of 0.5\,$\mu$m. In the high-speed system, a total of only 16 binarized images are taken with an interval of 4\,$\mu$m, which are insufficient to derive the precise three-dimensional position of each grain. 

\begin{figure}
\center
\includegraphics[bb=0 0 1455 741, width=.9\textwidth]{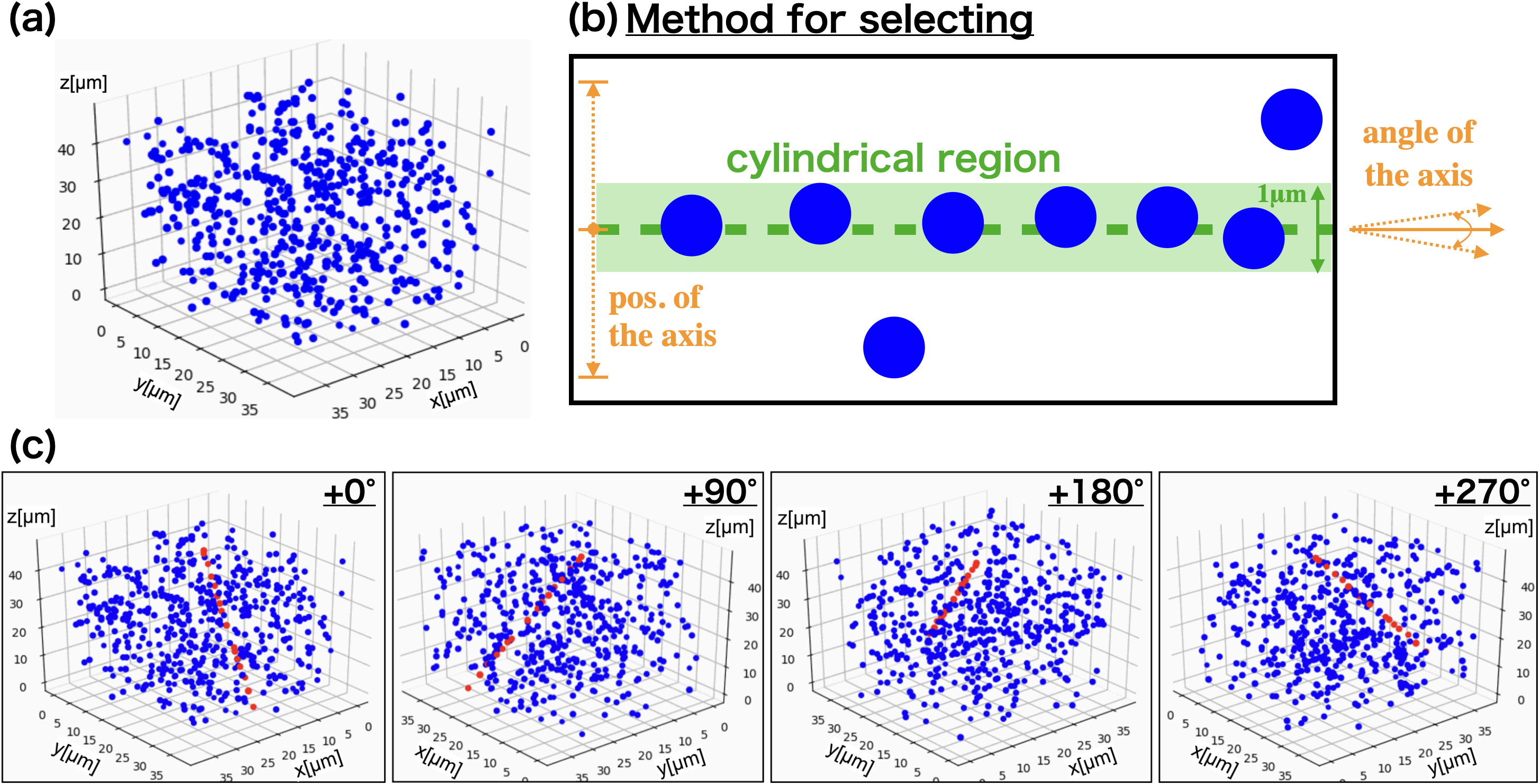}
\caption{(a): Example of three-dimensional positions of detected grains with the high-precision system (blue  dots). (b): Schematic illustration of the method for selecting the silver grains that are aligned along a straight line and composes the target-track. (c): Example of detected grains of a target-track (red dots) superposed in the same plot as in panel (a) in the left-most panel and in basically the same plot but viewed from three different angles rotated by an interval of 90$^\circ$ around the z-axis in the other panels.}
\label{fig:3DPos}
\center
\end{figure}

Here we detail the procedure of rescanning a track (hereafter referred to as "target-track'') once detected in the high-speed system, using the high-precision system. First, images in the vicinity of the target-track are taken by the high-precision system, using the XY-position obtained with the high-speed system, and the three-dimensional positions of the silver grains are measured. Figure \ref{fig:3DPos}(a) shows the three-dimensional positions of the detected silver grains. Next, a three-dimensional cylindrical region with a 1.0-$\mu$m diameter is defined in the three-dimensional space for the selection of the grains that composes the target-track. Figure \ref{fig:3DPos}(b) shows a schematic demonstration of the method to select the grains. The position and the angle of the axis of the cylindrical region are basically the position and the angle of the target-track obtained with the high-speed system (hereafter referred to as X,Y$_{\rm{target}}$ and tan$\theta_{\rm{target}}$). When the motor-driven stage move to the expected position of the target-track in the high-precision system, the position after moving may be slightly shifted from the position of the target-track because of the moving accuracy of the stage (<10\,$\mu$m). Considering this possibility, we shift the position of the axis of the cylindrical region by 1\,$\mu$m each for the X and Y directions within $-20$\,$\mu$m$<$X,Y$_{\rm{target}}$$<$20\,$\mu$m in searching for  the target-track. The angle of the target-track may vary, depending on the inclination of the film set on the stage, and expansion/contraction of the emulsion layer depends on humidity. For these reasons, we shift the angle of the axis of the cylindrical region by 0.005 each for the X and Y directions within -0.05$<$tan$\theta_{\rm{target}}$$<$0.05 in searching. We determine the silver grains of the target-track when the number of grains in the cylindrical region are maximum. Finally, we fit a straight line to the selected grains and decide the position and angle of the target-track in the emulsion layer in the high-precision system. Figure \ref{fig:3DPos}(c) shows detected grains of a target-track. We note that the high-precision system measures the precise parameters of low-energy particles, which the high-speed system cannot do well. Low energy particles often leave highly bent tracks due to multiple Coulomb scattering. The measurement of such bent tracks by the high-speed system, relying on the straight-line approximation in detection, inevitably suffers  a large angular uncertainty. The high-precision system directly measures the silver grains that composes the track regardless of its curvature and thus can measure even low-energy particles when we use other method to select the grains of the tracks of the low-energy particle. We estimated the improvement of the angular resolution for the low-energy gamma-ray with the measurement of each silver grain in a simulation, 1.0 $\rightarrow$ 0.6 $^\circ$ at 100\,MeV and 1.8 $\rightarrow$ 1.0 $^\circ$ at 60\,MeV \cite{GRAINE}.  This is another advantage of the high-precision system.

\subsection{Positional accuracy for each detected silver grain}
We evaluate the positional accuracy of our high-precision scanning system. A preliminary result was presented in our previous work\cite{GRAINE2018_2}. We select the high-energy tracks penetrating the converter from the top to bottom with negligible effect of multiple Coulomb scattering in GRAINE2018. We require that the standard deviation of the track angles detected in each of the stacked films is smaller than the angular resolution of the track detected with the high-speed system. The selection process with a set threshold filters out 99.9\% of 10\,GeV proton tracks; this filtering is hereafter referred to as the "high-energy filtering". As a result, each of these high-energy tracks consists of very straightly aligned silver grains in the emulsion layer, and the standard deviation of the position difference distribution between the position of each silver grain and the fitted straight line represents the positional accuracy for each silver grain detected with the high-precision system.

\begin{figure}
\center
\includegraphics[bb=0 0 1462 539, width=.99\textwidth]{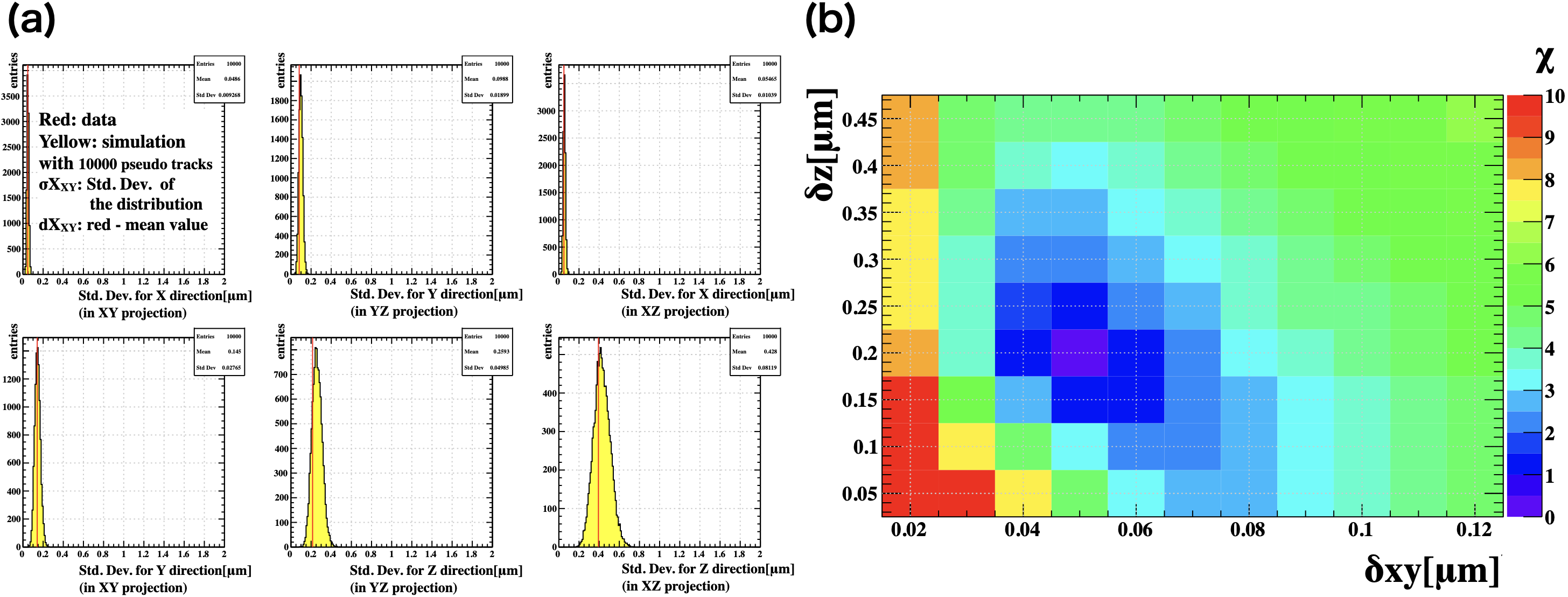}
\caption{(a) Red line in each panel shows the derived standard deviation of the position differences between the position of each silver grain and the fitted straight line for one real track. Yellow distribution of the standard deviation shows the standard deviations repeatedly derived for 10000 pseudo tracks with a provisional positional accuracy of each silver grain. Here, the provisional positional accuracy in the direction along the X-Y plane\ ($\delta xy$) is 0.05\,$\mu$m, and that in the Z direction\ ($\delta z$) is 0.25\,$\mu$m.  (b) Provisional positional accuracy dependence of the calculated probability parameter (see text for detail of the calculation).}
\label{fig:Dev}
\center
\end{figure}

\begin{figure}
\center
\includegraphics[bb=0 0 1271 830, width=.8\textwidth]{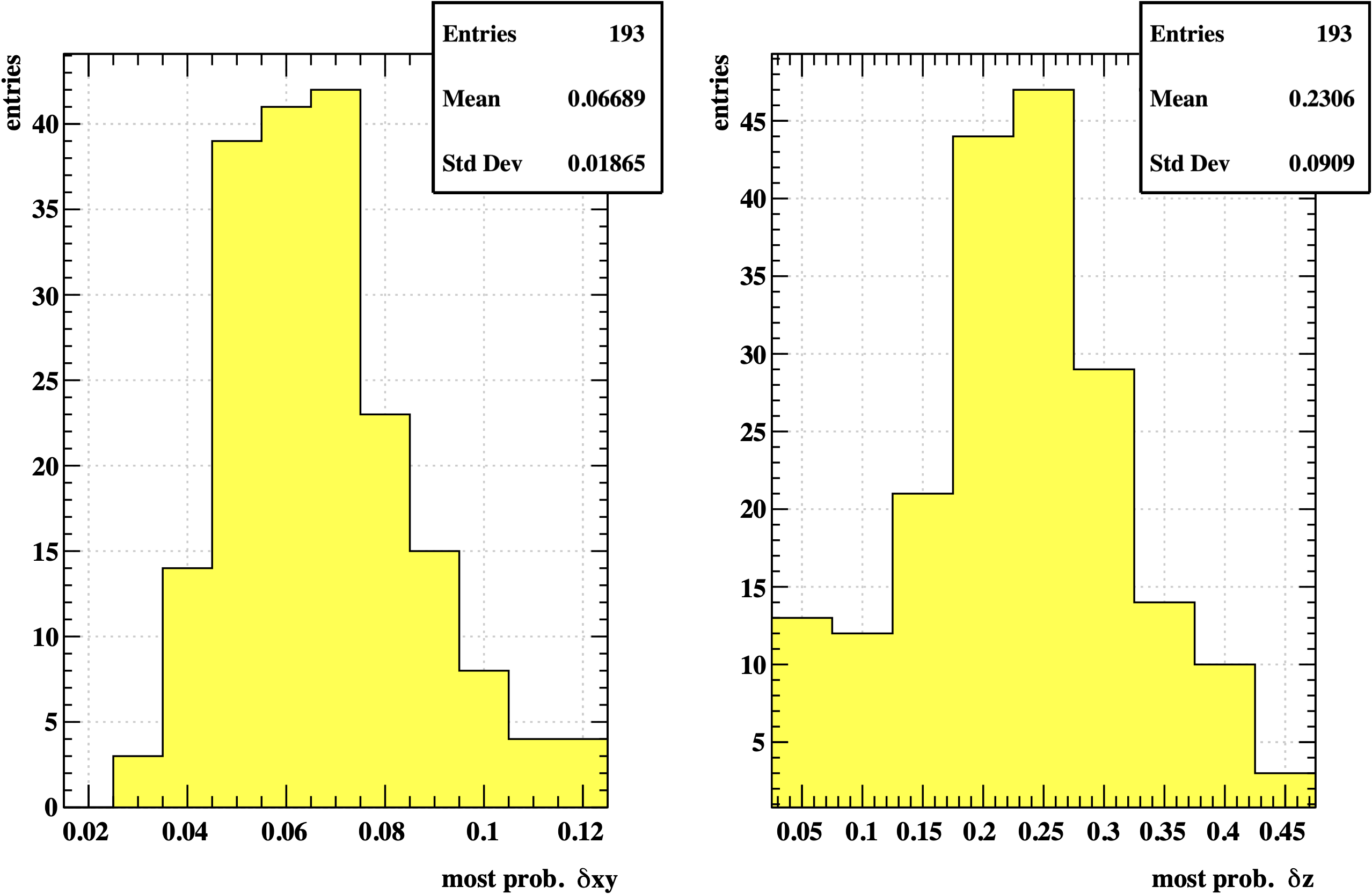}
\caption{Distribution of the most probable positional accuracy obtained from 193 real tracks and pseudo tracks. The mean values indicate the positional accuracy of each silver grain with the high-precision system in the direction along the X-Y plane, $\delta xy$ = 0.067 $\pm$ 0.001\,$\mu$m, and that in the Z direction, $\delta z$ = 0.231 $\pm$ 0.007\,$\mu$m.}
\label{fig:DevDist}
\center
\end{figure}

The standard deviation of the position difference depends on the positional accuracy and the incident angle of the track, and selected high-energy tracks have various incident angles. Considering it, we generate pseudo tracks using simulations and compare the position difference distribution with that of each real high-energy track for the evaluation of the positional accuracy. Figure \ref{fig:Dev}(a) shows an example of the obtained standard deviation for one real track with the high-energy filtering and simulated values for 10000 pseudo tracks. When we generate the pseudo tracks, we set that the incident angle and the number of silver grains of each pseudo track are the same as those of the real track, and the position of each silver grain of each pseudo track is randomly shifted from the straightly aligned position according to Gaussian distribution and a provisional positional accuracy that we assume. The mean value of the standard deviations obtained from the pseudo tracks indicate the expected standard deviation with the provisional positional accuracy. Then, we estimate the reasonable positional accuracy. We define $\chi$ as the probability parameter of the provisional positional accuracy. $\chi$ is calculated by the following equation:
\begin{equation}
\chi=\sqrt{(\frac{dX_{\rm{XY}}}{\sigma X_{\rm{XY}}})^2+(\frac{dY_{\rm{XY}}}{\sigma Y_{\rm{XY}}})^2+(\frac{dY_{\rm{YZ}}}{\sigma Y_{\rm{YZ}}})^2+(\frac{dZ_{\rm{YZ}}}{\sigma Z_{\rm{YZ}}})^2+(\frac{dX_{\rm{XZ}}}{\sigma X_{\rm{XZ}}})^2+(\frac{dZ_{\rm{XZ}}}{\sigma Z_{\rm{XZ}}})^2}
\end{equation}
where $dX_{\rm{XY}}$ is the difference between the standard deviation obtained from the real track and the mean value of the simulated values obtained from the pseudo tracks, and $\sigma X_{\rm{XY}}$ is the standard deviation of the simulated values; these are calculated for the X direction in the XY projection of the silver grains (see the top left in Figure \ref{fig:Dev}(a)). $dY_{\rm{XY}}$, $dY_{\rm{YZ}}$, $dZ_{\rm{YZ}}$, $dX_{\rm{XZ}}$, $dZ_{\rm{XZ}}$, $\sigma Y_{\rm{XY}}$, $\sigma Y_{\rm{YZ}}$, $\sigma Z_{\rm{YZ}}$, $\sigma X_{\rm{XZ}}$ and $\sigma Z_{\rm{XZ}}$ are same for the X,\ Y,\ and Z directions in the XY,\ YZ,\  and XZ projections, respectively. Figure \ref{fig:Dev}(b) shows the $\chi$ values when we set the different values as the provisional positional accuracy. In this example, the most probable positional accuracy in the direction along the X-Y plane\ ($\delta xy$) is 0.05\,$\mu$m and that in the Z direction\ ($\delta z$) is 0.20\,$\mu$m. Then, we evaluate the most probable positional accuracy with same method used for the 193 real tracks which is randomly selected from tracks with the high-energy filtering for the incident angles of 0.0 $<$ tan$\theta$ $<$ 1.0. The number of silver grains of the selected tracks is 10-20. Figure \ref{fig:DevDist} shows the result. The mean values, which indicate the accuracy of the high-precision system, are $\delta xy$ = 0.067$\pm$0.001\,$\mu$m and $\delta z$ = 0.231$\pm$0.007\,$\mu$m. These values are approximately an order of magnitude smaller than those with the latest high-speed system, the HTS ($\delta xy$ $\simeq$ 0.4\,$\mu$m, $\delta z$ $\simeq$ 2.0\,$\mu$m evaluated in GRAINE2018). The intrinsic accuracy is estimated from the diameter of AgBr crystal in the emulsion layer to be 240\,nm/$\sqrt{12} \simeq$ 69\,nm = 0.069\,$\mu$m. More realistically, the shape of the AgBr crystal is like a octahedron and the sensitivity for the charged particle is not uniform within the crystal, and the expected intrinsic accuracy is 0.060--0.069\,$\mu$m. Thus, the derived $\delta xy$ is consistent with the intrinsic value, whereas $\delta z$ is larger by a factor of three. Considering that the estimated $\delta z$ value may be affected by the depth of field of the objective lens and distance between each tomographic image, we conjecture that we could reduce $\delta z$ by shortening the distance.

\subsection{Connecting tracks in both emulsion layers}
\begin{figure}
\center
\includegraphics[bb=0 0 1259 750, width=.5\textwidth]{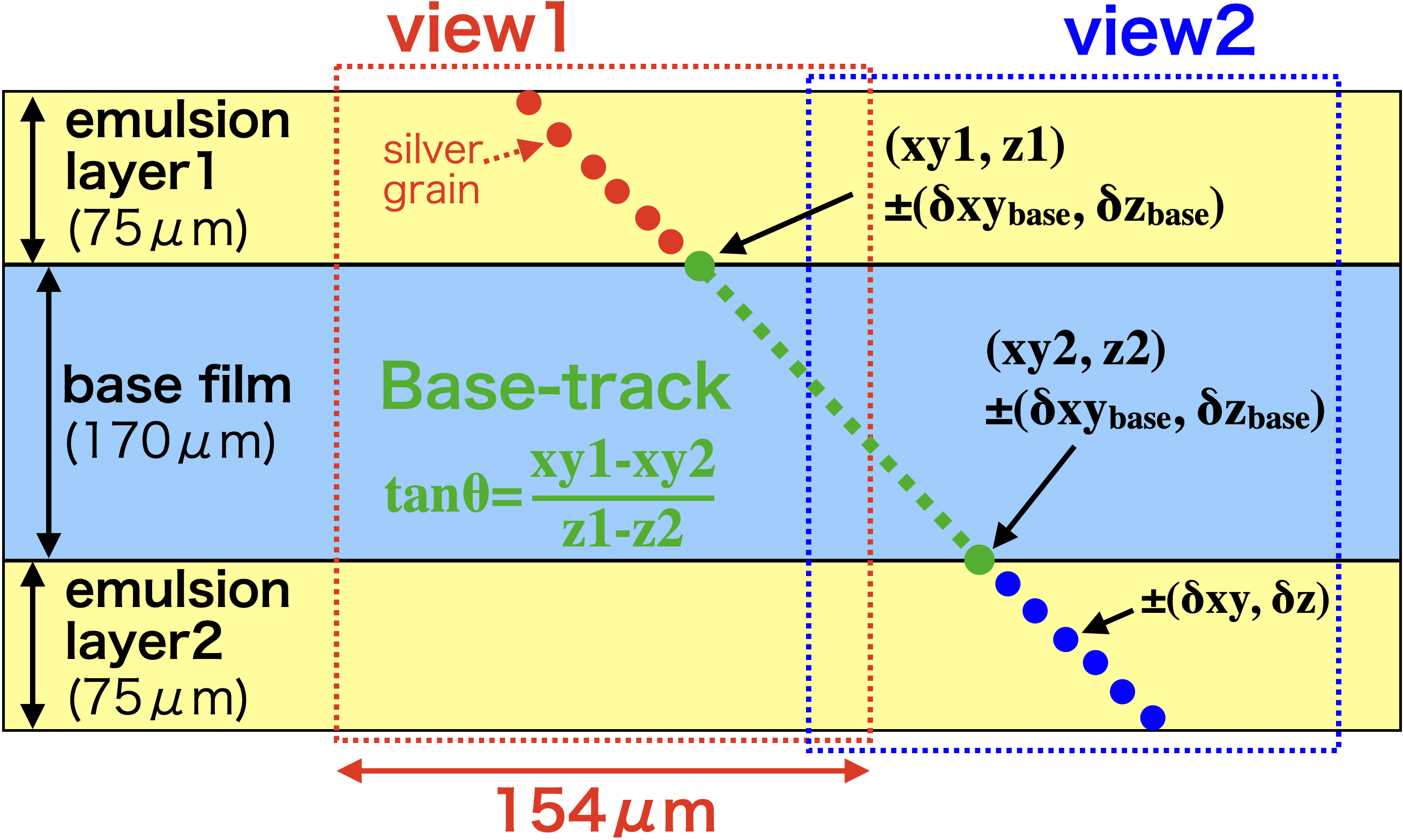}
\caption{Schematic drawing of the cross sectional view of the emulsion film and base-track. The positional resolutions of each silver grain in the direction along the X-Y plane and in the Z direction are referred to as $\delta xy$ and $\delta z$, respectively. The passage positional resolutions at one side of the surface of the base film are referred to as $\delta xy_{\rm{base}}$ and $\delta z_{\rm{base}}$.}
\label{fig:basetrack}
\center
\end{figure}

In the emulsion film, the plastic base-film layer is sandwiched with emulsion layers (Figure~\ref{fig:proc}). When a target-track is detected in both the emulsion layers, a virtual track in the base film, hereafter referred to as "base-track", is defined, which is a guessed passage inside the base film linking the tracks in both the emulsion layers (see Figure \ref{fig:basetrack}). The base-track is not affected expansion/contraction of the emulsion layers, and its angular resolution is better than the tracks in the emulsion layers because the base film is thicker than an emulsion layer. For these reasons, the base-track is a useful concept for analysis. The tracks in the emulsion layers are used with a correction of the expansion/contraction of the emulsion layers through comparison of the angle of the base-track and the angles of the tracks in the emulsion layers when we want to use the tracks for analysis.

The high-precision system is not able to take tomographic images of the tracks in both the emulsion layers with large incident angles in one view, and two or three views are used to make base-track (see Figure \ref{fig:basetrack}). The relative position between the two views may slightly deviate from the true value due to the moving accuracy of the motor-driven stage. Considering this possibility, we correct the relative position with the positions of the detected silver grains in the overlapping region of the two views. An aberration of the objective lens affecting the positions of the silver grains detected especially in the edge region of the view, including the overlap region, is corrected by measuring length of a stage micrometer. Then, we evaluate the correction accuracy of the relative position between the two views by comparing the correction values calculated individually in each layer, and the accuracy is less than 70\,nm.

\subsection{Angular resolution for high-energy proton beams}
The angular resolution of the base-track with the high-precision system is evaluated with beam experiments, using 400-GeV proton beams with a very good parallelity in Super Proton Synchrotron at CERN. The beam has an angular size of 38\,$\mu$rad, which is enough smaller than the intrinsic angular resolution of the base-track ($\sim$1\,mrad). Thus, the standard deviation of the detected angles ($\sigma$tan$\theta$) represents the angular measurement accuracy of the scanning system. The incident angles of the proton beams are varied with a step of tan$\theta_{\rm{X}}\sim$0.5 for a range of $-$2.0 $<$ tan$\theta_{\rm{X}}$ $<$ 2.0 with a rotary table while tan$\theta_{\rm{Y}}\sim$0.0 is maintained throughout. Further details in the experimental setup are presented in \cite{CERN}.

\begin{figure}
\center
\includegraphics[bb=0 0 1271 820, width=.8\textwidth]{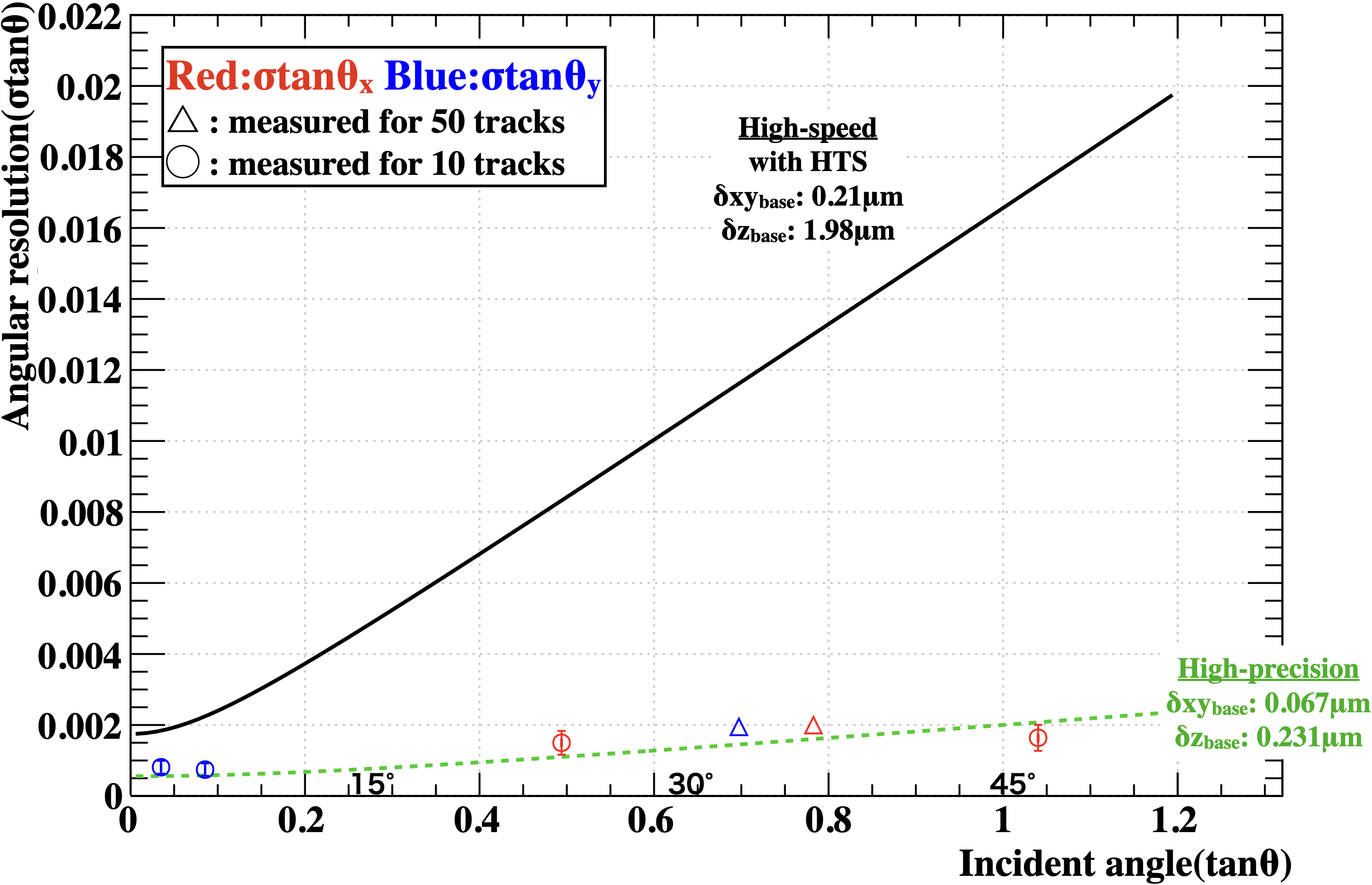}
\caption{Angular resolution in the observation of the base-track generated by incoming high-energy charged particle. Red and blue open circles show our experimental results evaluated for 10 tracks in exposed incident angles of tan$\theta_{\rm{X,Y}}\sim0.5,0.0$ and tan$\theta_{\rm{X,Y}}\sim1.0,0.0$ for the X-axis and Y-axis direction, respectively. Open triangles show those for 50 tracks in angles of tan$\theta_{\rm{X,Y}}\sim0.7,0.7$, where the emulsion film on the motor-driven stage is arranged to positioned with a rotation angle of $\sim$45$^\circ$. Black and green dashed lines show the calculated value with parameters given in the figure (see text for detail of the calculation).}
\label{fig:PreBtAngRes}
\center
\end{figure}

Figure \ref{fig:PreBtAngRes} presents the incident angle dependence of the evaluated angular resolution of the base-track with the high-precision system. Here, we calculate the expected angular resolution of the base-track($\delta$tan$\theta$) according to:
\begin{equation}\label{eq:AngRes}
\delta tan\theta=\frac{\sqrt{2}}{170}\sqrt{\delta xy_{\rm{base}}{^2+(\delta z_{\rm{base}}\cdot tan\theta)^2}},
\end{equation}
where the denominator 170\,($\mu$m) means the thickness of the base film in analysis. Although the real thickness of the base film is 180\, $\mu$m, the scanning system recognize the thickness as $\sim$170\, $\mu$m because of the difference of the refractive indices between the emulsion layer and base film. $\delta xy_{\rm{base}}$ and $\delta z_{\rm{base}}$ are the passage positional resolutions at one side of the surface of the base film in the direction along the X-Y plane and in the Z direction, respectively. This result shows that the angular resolution with the high-precision system is significantly improved than that with the high-speed system, $\delta$tan$\theta$: 0.0022 $\rightarrow$ 0.0006$\pm$0.0002 for tan$\theta\sim$0.08, and especially for a large incident angle, for which about one order of magnitude improvement ($\delta$tan$\theta$: 0.017 $\rightarrow$ 0.0016$\pm$0.0004) is observed for tan$\theta\sim$1.04. The calculated angular resolution (0.0021 for tan$\theta\sim$1.04 and 0.0006 for tan$\theta\sim$0.08) is consistent with the experimental results (0.0016$\pm$0.0004 for tan$\theta\sim$1.04 and 0.0007$\pm$0.0002 for tan$\theta\sim$0.08) in the case where $\delta xy_{\rm{base}}$ and $\delta z_{\rm{base}}$ are equal to $\delta xy$ and $\delta z$ (evaluated in Section 3.3), respectively. The passage positions are determined with straight-line approximation of some silver grains, and its uncertainty ($\delta xy_{\rm{base}}$) may be better than the positional accuracy for each silver grain ($\delta xy$). The value of $\delta xy_{\rm{base}}$ with the high-precision system may affect the position-correction accuracy (see the previous section), and because of this, $\delta xy_{\rm{base}}$ is not better than $\delta xy$. 

\subsection{Coordinate transformation from the high-precision system to the high-speed system}
Flatness of the films in the converter and the measured incident angles of the gamma-rays are corrected with the alignment films (described in Section 2), using tracks affected negligible multiple Coulomb scattering. These tracks recorded in the converter and the alignment films should be measured with the high-speed system, as opposed to the high-precision system, because many tracks are needed for the correction for many stacked films and for a large area in the converter. The corrected gamma-ray events in the high-speed system coordinates are mapped to celestial coordinates with using the relative angle between the alignment films and attitude monitor. Thus, we can not directly correct the track angles measured with the high-precision system in the converter to the alignment films (and attitude monitor) coordinates, and we need to convert the track angles measured with the high-precision system to the angles with the high-speed system coordinates for astronomical observation. 

In practice, 100 tracks detected with the high-speed system in the vicinity of the target-track are read out with the high-precision system for the coordinate transformation. Hereafter, these tracks are referred to as "transform-tracks". The transform-tracks are randomly selected from the tracks with the high-energy filtering (described in Section 3.3). The incident angles\ (tan$\theta$) of 30 tracks among the 100 tracks measured with the high-speed system are distributed in 0 $<$ tan$\theta$ $<$ 0.2, and those of the other 70 tracks are in 0.7 $<$ tan$\theta$ $<$ 1.0, presumably because a track with a larger incident angle has a larger angular uncertainty, especially in the high-speed system. Then, the transformation parameters in the angular parameter space with rotation, expansion/contraction, and offsets are obtained through fitting and application of the following equation: 
\begin{equation}\label{eq:affin}
\begin{bmatrix}\mathrm{tan}\theta_{\rm{X}}' \\\mathrm{tan}\theta_{\rm{Y}}' \end{bmatrix} = r\times(\begin{bmatrix}\mathrm{cos\phi}~~\mathrm{sin\phi} \\\mathrm{-sin\phi}~~\mathrm{cos\phi} \end{bmatrix} \begin{bmatrix}\mathrm{tan}\theta_{\rm{X}} \\\mathrm{tan}\theta_{\rm{Y}} \end{bmatrix} + \begin{bmatrix}p  \\q  \end{bmatrix}),
\end{equation}
where tan$\theta_{\rm{X,Y}}$ and tan$\theta_{\rm{X,Y}}'$ are the base-track angles before and after the correction, respectively, $r$ is the factor of expansion/contraction, $\phi$ represents that of rotation, and $p$ and $q$ are offsets for the X and Y directions, respectively.

\begin{figure}
\center
\includegraphics[bb=0 0 1265 828, width=.8\textwidth]{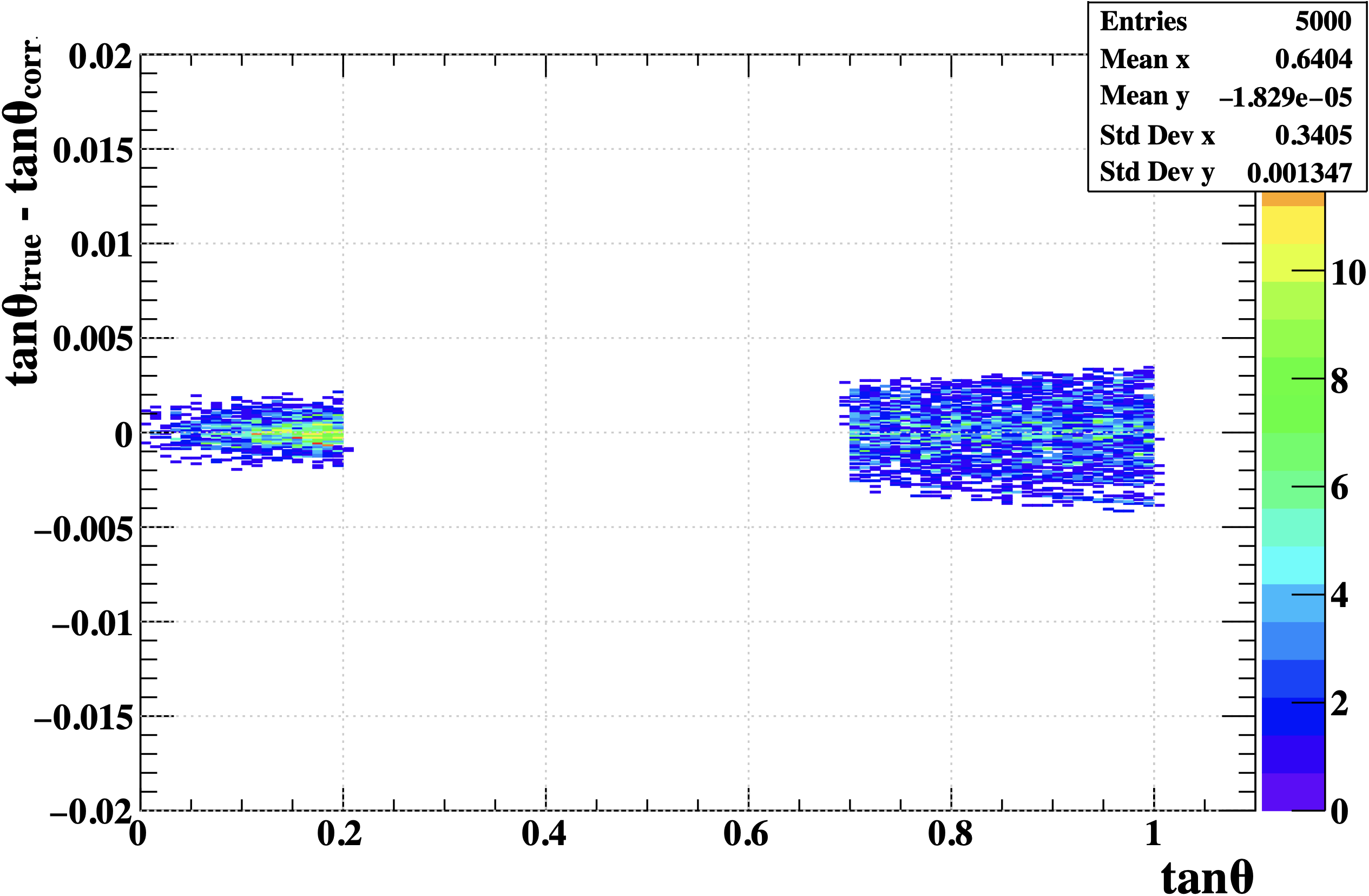}
\caption{Angule-dependence of the angle-difference between the simulated model angle and reconstructed one in the simulation with the coordinate transformation method (see text). 100 pseudo tracks are used in each of 50 trials for the results.}
\label{fig:SimTransAcc1D}
\center
\end{figure}

\begin{figure}
\center
\includegraphics[bb=0 0 1178 841, width=.8\textwidth]{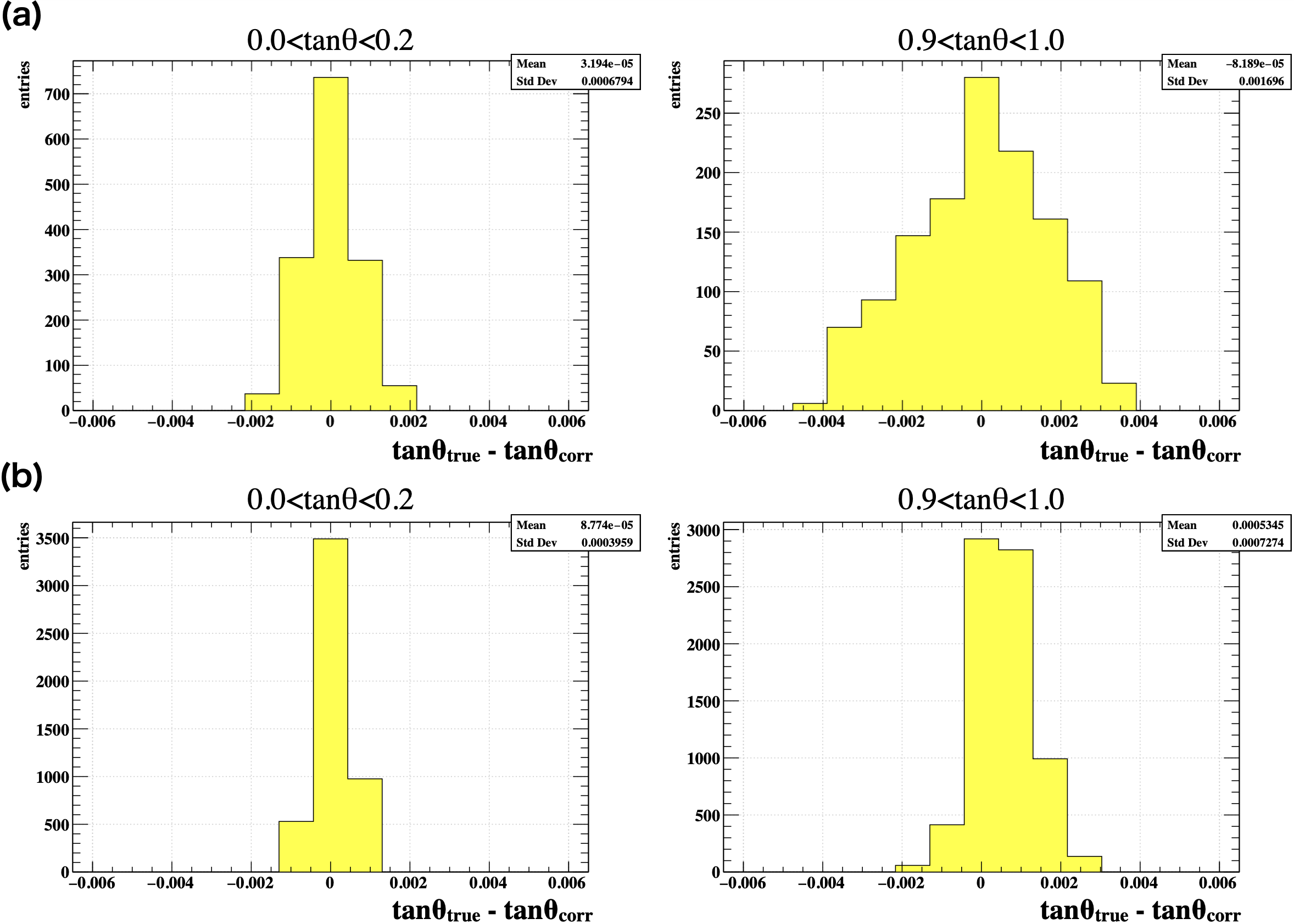}
\caption{Angle-difference distribution between the simulated model angle and reconstructed one in the simulation with the coordinate transformation method (see text) for the different angular region (0 $<$ tan$\theta$ $<$ 0.2 and 0.0 $<$ tan$\theta$ $<$ 1.0). In the derivations,100 and 500 pseudo tracks are used in each of 50 trials for the results in panels (a) and (b), respectively.}
\label{fig:SimTransAcc}
\center
\end{figure}

In the coordinate transformation, the angular resolution with the high-precision system should be maintained. We evaluate the accuracy of this transformation method with Monte Carlo simulations. Firstly, 100 pseudo transform-tracks are generated In this simulation and the incident angles of these tracks are treated as tan$\theta$ in Equation \ref{eq:affin}, and tan$\theta'$ is calculated according to the equation with assumed transformation parameters of $r$=1.1, $\phi$=0.04\,rad and $p$=0.08,\ $q$=0.05, all of which are same as the worst-case of the realistic values as conservative trials. Next, each of the track angles before the correction (tan$\theta$) is randomly shifted according to the angular resolution with the HTS in Figure \ref{fig:PreBtAngRes} and Gaussian distribution, and each of them after the correction (tan$\theta'$) is randomly shifted according to that with the high-precision system and Gaussian distribution. Then, we derive the transformation parameters by combining the obtained angles after the randomly shifting with Equation \ref{eq:affin}. Finally, we evaluate the transformation accuracy through comparison of the simulated model angle with the calculated angle in combination with the estimated transformation parameters. Figure \ref{fig:SimTransAcc1D} shows the angle-dependence of the angle-difference obtained with 50 trials. Figure \ref{fig:SimTransAcc}(a) shows the results of the evaluation of the transformation accuracy, and it shows the distribution of the angle differences for the different angular region. The standard deviation of this distribution, which represents the angular transformation accuracy, is $\sim$0.0007 in 0 $<$ tan$\theta$ $<$ 0.2 and $\sim$0.0017 in 0.9 $<$ tan$\theta$ $<$ 1.0. Figure \ref{fig:SimTransAcc}(b) shows a similar plot but with 500 pseudo transform-tracks (100 tracks are distributed in 0 $<$ tan$\theta$ $<$ 0.2, and other 400 tracks are in 0.7 $<$ tan$\theta$ $<$ 1.0) and 50 trials;  the standard deviation is  $\sim$0.0004 and $\sim$0.0007. These results indicate that the number of the transform-tracks should be set according to the desired angular resolution, and that the transformation accuracy is better than the angular resolution of the base-track in the high-precision system (Figure \ref{fig:PreBtAngRes}) if 500 transform-tracks are used. We need more time to scan the number of more transform-tracks, and we set the reasonable number of that in the gamma-ray analysis according to the expected angular resolution depends on the multiple Coulomb scattering.


\section{Performance with gamma-rays with the high-precision system}
\subsection{Evaluation method}
We evaluate the performance of our newly developed high-precision system with gamma-ray events detected in GRAINE2018. The gamma-ray directions and energies were already once reconstructed  with the HTS, where the angles and the momenta of the electron and positron tracks were used for the reconstruction (see \cite{GRAINE2018_2}\cite{GRAINE_gamma} for detail). In this performance evaluation, we rescan the positron and electron tracks in the films with the high-precision system and determine the angles, whereas we use the already derived momenta with the high-speed system\cite{GRAINE2018_2} for the gamma-ray reconstruction.

\begin{figure}
\center
\includegraphics[bb=0 0 860 840, width=.6\textwidth]{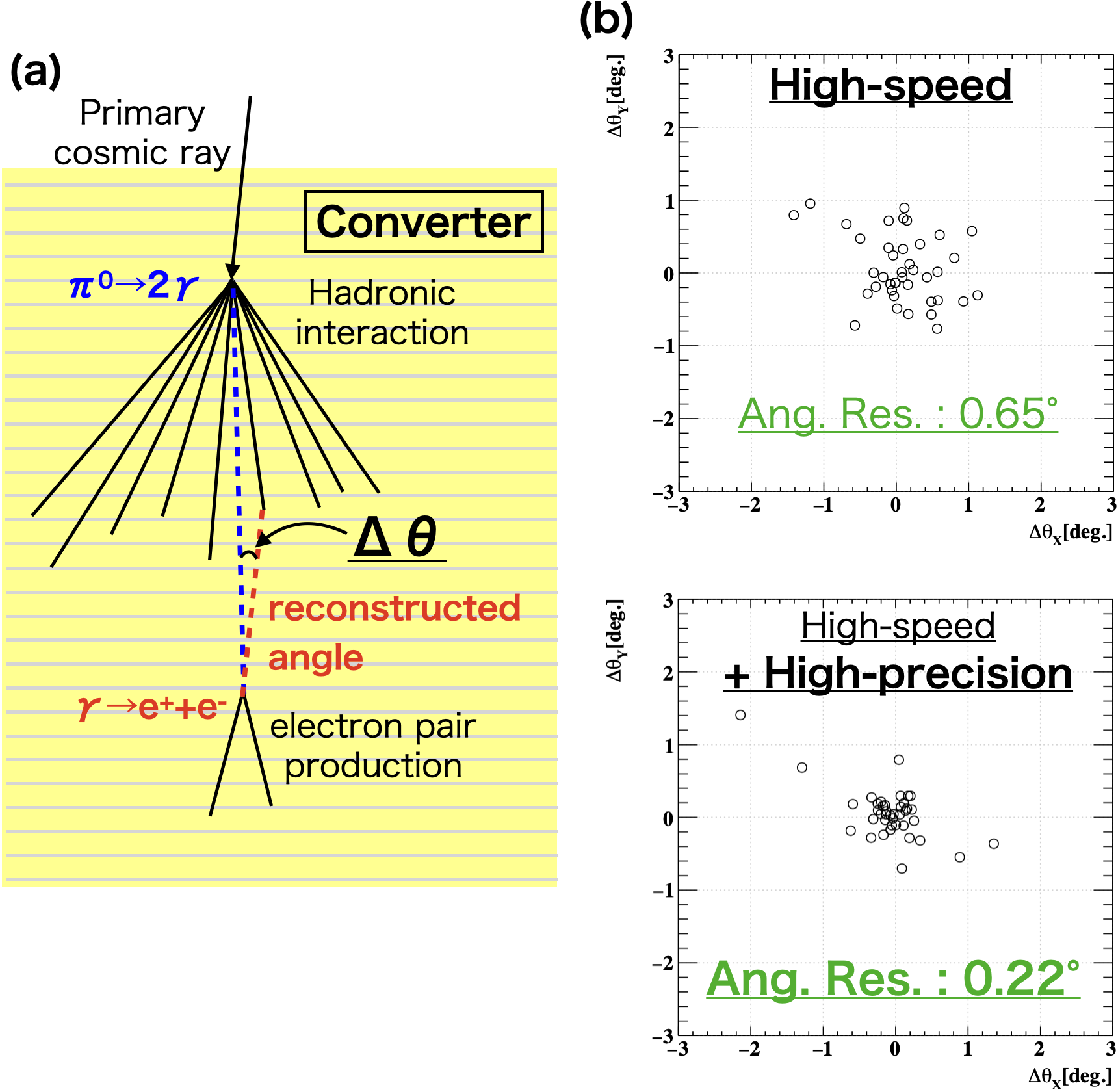}
\caption{(a) Evaluation method of the gamma-ray observation performance with the flight data. (b) Distribution of the angle differences, $\Delta\theta$, obtained with (top panel) the high-speed system alone and (bottom panel) a combination of our newly developed high-precision system with the high-speed system.}
\label{fig:Gamma}
\center
\end{figure}

Figure \ref{fig:Gamma}(a) illustrates our evaluation method of angular resolution of detected gamma-rays. We used concomitant gamma-rays, i.e., gamma-rays produced in $\pi^0$ decays\ ($\pi^0 \rightarrow 2\gamma$), after $\pi^0$ are produced in hadronic interactions of cosmic rays (such as, protons and helium nucleus) in the converter during observation. $\pi^0$ has a lifetime of only $\sim$ 80\,attoseconds. Thus, we treat the interaction points of the hadronic interactions as the decay positions of $\pi^0$, and the arrival direction of these gamma-rays are expected by connecting the hadronic interaction points and the conversion points of the pair-productions from the gamma-rays.  We can evaluate the angular resolution of the reconstructed gamma-rays with the distribution of the angle difference, $\Delta\theta$, between the expected arrival direction and the reconstructed incoming directions of gamma-rays. This evaluation method was used with GRAINE2018\cite{GRAINE2018_2}. In this study, we reanalyze the randomly selected 40 concomitant gamma-ray events in 500--700\,MeV and for 0.8 $<$ tan$\theta$ $<$ 1.0 with the high-precision system. The 40 events are almost all available events in this energy and angular range. The 40 samples are statistically expected to contain two background events, i.e., those not from $\pi^0$ decays. Also, 100 transform-tracks within 0.5\,cm for the XY direction from each gamma-ray event are rescanned with the high-precision system; the transformation accuracy with 100 transform-tracks is enough smaller than the expected angular resolution considering the multiple Coulomb scattering in this energy range. More than 5000 tracks remain within the region in GRAINE2018 after applying the high-energy filtering (described in Section 3.3), and we randomly selected 100 tracks among them as the transform-tracks.

\subsection{Results}
Figure \ref{fig:Gamma}(b) shows the distribution of the angle differences, $\Delta\theta$, obtained with the high-speed system alone and a combination of our newly developed high-precision system with the high-speed system. The origin of each distribution is the expected arrival direction. The incoming directions of gamma rays obtained with the reanalysis with the high-precision system are found to be more concentrated in the vicinity of the origin. There are a few events which presents instead a much larger angle-difference, and the number of this outlier events are compatible with the expected number of the background events (described in the previous section). The angular resolutions are calculated to be $0.65^{+0.09}_{-0.19}$ $^\circ$ and $0.22^{+0.22}_{-0.09}$ $^\circ$ for the former and latter configurations, respectively, each of which is defined as the radius of a circle that contains 68$\%$ of the events after subtraction of the uncertainty of the expected direction calculated for each incidence-angle and energy ranges (see \cite{GRAINE2018_2}). Each error represents the statistical uncertainty. Thus, the angular resolution obtained with the latter, primarily the high-precision system, is about three times smaller than that obtained with the high-speed system.

\begin{figure}
\center
\includegraphics[bb=0 0 1266 834, width=.8\textwidth]{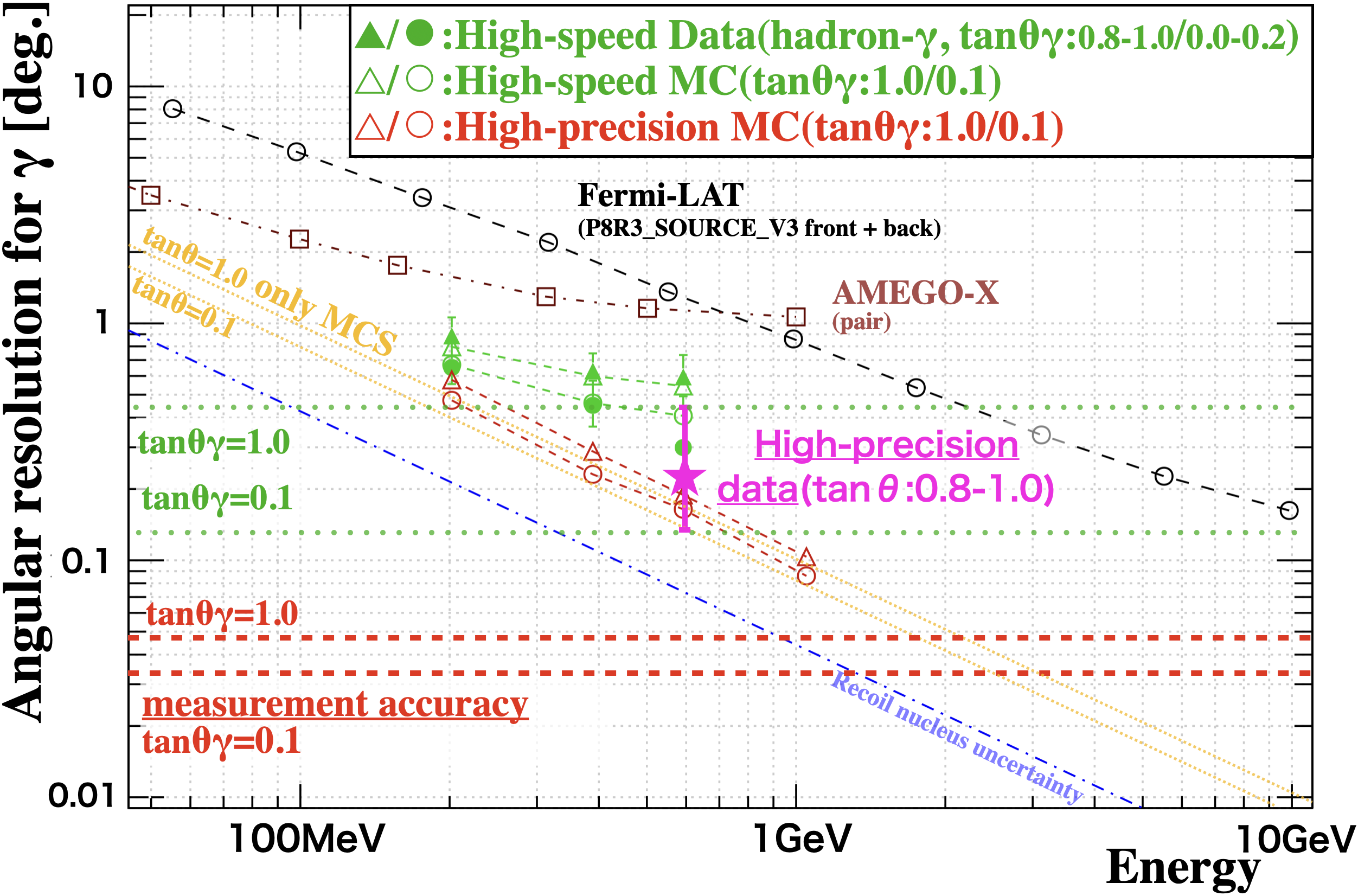}
\caption{Energy dependence of the angular resolution for gamma rays. Yellow solid lines are the simulated angular resolutions, where only multiple Coulomb scattering is considered. Green-dotted and red-dashed horizontal lines are the angular measurement accuracy with the high-speed system and our newly developed high-precision system, respectively, for two incident angles. Green open triangles and circles are Monte-Carlo (MC) simulated resultant angular resolutions and those derived with the high-speed system, respectively, for gamma-ray events for two incident angles. Red markers are the same as that but with the high-precision system. Green filled triangles and circles are evaluated values with the high-speed system based on the flight data in our previous work\cite{GRAINE2018_2}. The magenta star shows the evaluated value in this study obtained with the high-precision system, partially combined with the data of the high-speed system. Black and brown markers and lines show the angular resolution for gamma-rays with the current and future satellite mission, Fermi-LAT and AMEGO, respectively\cite{FermiAngRes}\cite{FermiPerformance}\cite{AMEGOAngRes}.}
\label{fig:GammaAngRes}
\center
\end{figure}

Figure \ref{fig:GammaAngRes} shows the obtained and simulated angular resolutions as a function of energy. We calculate the intrinsic angular resolution for gamma-rays through a Monte Carlo simulation using the Geant 4.10.01, where only multiple Coulomb scattering of the electron-positron pairs in the emulsion film is considered\cite{geant}. We evaluated the resultant angular resolution with the effect of multiple Coulomb scattering and the measurement accuracy with the high-speed system based on the flight data in our previous work\cite{GRAINE2018_2}, and the resultant resolution is 1.5--4 times larger than the intrinsic resolution in 100--700\,MeV band due to the insufficient measurement accuracy of the high-speed system. By contrast, the newly developed high-precision system achieved 4--9 times higher measurement accuracy for gamma-rays than the high-speed system, and the expected resultant resolution is almost the same as the intrinsic resolution with an achieved resolution of $\sim$0.1$^\circ$ at 1\,GeV. Furthermore, the derived angular resolution in this study based on the flight data (Figure~\ref{fig:Gamma}(b)) is found to be consistent with the expected one (Figure~\ref{fig:GammaAngRes}).

\section{GRAINE2023 experiment and the planning of applying of the high-precision system}
The fourth balloon experiment, GRAINE2023, was conducted in 2023 April in Australia for a total flight duration of 27 hours. Its aperture area was 2.5\,m$^2$, 6 times larger than that in GRAINE2018. GRAINE2023 is the first step for scientific observation in the GRAINE project. The Galactic center region was observed for the first time with the emulsion gamma-ray telescope, the flight duration in GRAINE2018 was insufficient to observe the region. 

Although GRAINE2018 successfully observed the Vela pulsar with the record-high angular resolution for sub-GeV energy band, various factors significantly limited its performance; the primary factor was  the measurement accuracy with the high-speed scanning system as demonstrated in this paper, but there were also other significant factors, including azimuthal rotation of the gondola within the time resolution and the correction accuracy of the alignment films \cite{GRAINE2018_1}. Because of these factors, even if the new high-precision system is employed in GRAINE2018, the gained benefit for the angular resolution might be limited in observations for cosmic sources. By contrast, in GRAINE2023, the components that caused the other factors to degrade the angular resolution were updated from GRAINE2018 to the extent their uncertainty became negligible for the goal resolution of 0.1$^\circ$. Consequently, the measurement accuracy of the scanning system was more significant than ever to determine the observation performance in GRAINE2023, especially in the high-energy band.

The total analysis speed of gamma-ray events in GRAINE experiment almost limited by the scanning system. Then, the analysis speed of the new high-precision system for gamma-ray events is almost the same as that of the HTS alone. This is because it takes advantage of the speedy processing by the existing high-speed system; specifically, the new system delegates to the high-speed system the job of searching for gamma-ray events from a large number of tracks and reanalyzes only the selected events in order to achieve higher precision in parallel to the delegated searches for more events. A new high-speed system, called the HTS2, is being developed now, and its scanning speed will be more than 2 times faster than that of the HTS once manufactured. The processing speed of the current high-precision system may limit the total analysis speed. However, the new high-precision system being still a prototype, its processing speed may become more than 4 times faster with some of the hardware updated to better ones; then, the processing speed of the updated high-precision system would become almost irrelevant to the total analysis speed.

The flight films in GRAINE2023 were scanned by the HTS and HTS2 first, and the data analysis is ongoing. We will apply the new high-precision scanning system to the data and achieve the observation analysis of the Vela pulsar and Galactic center region with the highest angular resolution and lowest contamination of the diffuse gamma-ray background. In addition, we plan to start polarization measurements of the Vela pulsar in the sub-GeV band although the expected number of events from it may not yet be sufficient for positive detection in GRAINE2023. Whereas we verified the polarization sensitivity of the emulsion film for the high-energy gamma-rays through the beam experiment \cite{GRAINE_polar} by eye using microscope with the intrinsic positional resolution of the emulsion film, we spent much time for the analysis and it was difficult to apply the polarization measurement to the astronomical observation. By contrast, the high-precision system realizes the measurement for electron pair tracks in a short time with the almost same positional resolution as the intrinsic one (we described in Sec. 3.3), and it will realize the polarization measurement of high-energy gamma-ray from astronomical sources for the first time.

\section{Summary and prospects}
We have been running the GRAINE project,  a cosmic gamma-ray observation project in an energy range of 10\,MeV -- 100\,GeV, using nuclear emulsion, characterized with a high-angular resolution, polarization sensitivity and large aperture area \cite{GRAINE}. In the third balloon experiment GRAINE2018, the emulsion gamma-ray telescope achieved the highest angular resolution ever reported in the energy range $>$80\,MeV \cite{GRAINE2018_1}\cite{GRAINE2018_2}. However, the measurement accuracy of the emulsion scanning system of GRAINE2018 was not sufficient to observe high energy gamma-rays (especially in the energy range $>$1\, GeV) with the intrinsic angular resolution of the emulsion film. For example, it should be improved to realize a high-resolution observation in the few GeV energy range to possibly disentangle the source emission of the unexpected GeV gamma-ray excess in the Galactic center region.

In this study, we developed a new high-precision scanning system. The concept is to make a rescan with the new high-precision system in the vicinity of only the gamma-ray events already detected with the high-speed system. The high-speed system reduce a huge number of the recorded track data to a manageable size in a speedy manner, and the new high-precision system with high measurement accuracy improve the observation performance. We developed a new high-precision measurement algorithm that  precisely identifies each silver grain that composes a track with a positional accuracy of $\delta xy$ = 0.067$\pm$0.001\,$\mu$m for the direction along the plane parallel on the film (X-Y plane) and $\delta z$ = 0.231$\pm$0.007\,$\mu$m for the direction perpendicular to it (Z direction). These are approximately an order of magnitude smaller than those obtained by the latest high-speed system ($\delta xy$ $\simeq$ 0.4\,$\mu$m, $\delta z$ $\simeq$ 2.0\,$\mu$m evaluated in GRAINE2018), and $\delta xy$  is almost the same as the intrinsic accuracy of the nuclear emulsion used in GRAINE. The angular resolution for a charged-particle track with the high-precision system was evaluated, using the film of a 400-GeV proton beam experiment, and the evaluated value was 5--10 times smaller than that with the latest high-speed system, depending on the incident angle. These results indicate that this new high-precision system is highly useful for not only gamma-ray observations but also other fields requiring charged particle measurements. Furthermore, we built an algorithm to combine the existing high-speed system with the new system without sacrificing the high angular resolution achieved with the new system alone.

Finally, we applied the new high-precision system to events from concomitant gamma-rays from hadronic interactions taken in GRAINE2018 and obtained the angular resolution of 0.22$^\circ$ in 500--700\,MeV and for 0.8 $<$ tan$\theta$ $<$ 1.0. This angular resolution is approximately three times smaller than that obtained with the latest high-speed system. Simulations with the measurement accuracy of the new system show that the obtained value is consistent with the simulated one and also predict an angular resolution of $\sim$0.1$^\circ$ at 1\,GeV with this system. This resolution is better than that of near-future satellite missions, for example AMEGO, in the sub-GeV/GeV range. Furthermore, the high-precision system can measure the azimuthal angle of electron and positron pairs at less than $\sim$20\,$\mu$m below the conversion point. The fact demonstrates that high-positional-precision measurements of pair-production events will enable us to make the first polarization detection from cosmic sources in this energy range in the future. Polarization measurements would be highly useful to investigate the particle acceleration mechanism in pulsars, blazars, and so on.

We conducted fourth balloon experiment, GRAINE2023, in 2023 April in Australia. The flight succeeded to cover the duration to observe target sources, the Vela pulsar and the Galactic center. We aim to realize the observation with the highest angular resolution for the Vela pulsar and the Galactic Center in GeV band, and to realize the polarization measurement for the Vela pulsar in sub-GeV band. Gamma-ray radiation in the immediate vicinity of the Galactic center may bring new insights into the dark matter density profile among other exciting discoveries, and the polarization measurement may bring new insights into the particle acceleration mechanism, whereas the expected number of events may not yet be sufficient for positive detection in GRAINE2023. The high positional/angular resolution is important to realize such observations, and the emulsion film and the new high-precision system make it possible. Then, scientific observations in the GRAINE project started from GRAINE2023, while we will perform more balloon experiments with the high-precision system.

\section*{acknowledgments}
The balloon-borne experiment was conducted by the Scientific Ballooning (DAIKIKYU) Research and Operation Group, ISAS, JAXA (PI support: C. Ikeda) with CSIRO. English proofreading of this manuscript was done by a professional editor (Dr. Masaaki Sakano in Wise Babel Ltd). This work was supported by JSPS KAKENHI (grant numbers 17H06132, 18H01228, 18K13562 and 22K20382).

\end{document}